\documentclass[11pt]{article}
\usepackage{latexsym}
\usepackage{curves}
\newcommand{\be}{\begin{equation}}
\newcommand{\ee}{\end{equation}\noindent}
\newcommand{\bear}{\begin{eqnarray}}
\newcommand{\ear}{\end{eqnarray}\noindent}
\newcommand{\no}{\noindent}
\date{}
\renewcommand{\theequation}{\arabic{section}.\arabic{equation}}

\newcommand{\slD}{\raise.15ex\hbox{$/$}\kern-.57em\hbox{$D$}}
\newcommand{\slpartial}{\raise.15ex\hbox{$/$}\kern-.57em\hbox{$\partial$}}
\newcommand{\slG}{{{\dot G}\!\!\!\! \raise.15ex\hbox {/}}}

\def\non{\nonumber}
\def\beqn*{\begin{eqnarray*}}
\def\eqn*{\end{eqnarray*}}
\def\sy{\scriptscriptstyle}

\def\square{\kern1pt\vbox{\hrule height 1.2pt\hbox{\vrule width 1.2pt
   \hskip 3pt\vbox{\vskip 6pt}\hskip 3pt\vrule width 0.6pt}
   \hrule height 0.6pt}\kern1pt}
\def\slash#1{#1\!\!\!\raise.15ex\hbox {/}}

\def\dps{\displaystyle}
\def\sy{\scriptscriptstyle}
\def\half{{1\over 2}}

\def\fourth{{1\over4}}

\def\4piTD{{(4\pi T)}^{-{D\over 2}}}
\def\4piT4{{(4\pi T)}^{-2}}

\def\Tintm4{{\dps\int_{0}^{\infty}}{dT\over T}\,e^{-m^2T}
    {(4\pi T)}^{-2}}
\def\Tintm{{\dps\int_{0}^{\infty}}{dT\over T}\,e^{-m^2T}}

\def\be{\begin{equation}}\def\ee{\end{equation}}
\def\bea{\begin{eqnarray}}\def\eea{\end{eqnarray}}
\def\ba{\begin{array}}\def\ea{\end{array}}
\def\bea{\begin{eqnarray}}\def\barr{\begin{array}}\def\earr{\end{array}}
\def\eea{\end{eqnarray}}
\begin{document} \input feynman
\newcommand{\ho}[1]{$\, ^{#1}$}
\newcommand{\hoch}[1]{$\, ^{#1}$}
\newcommand{\kings}{\it\small Department of Mathematics, King's College, London, UK}
\newcommand{\lapp}{\it\small LAPP, Annecy, France}
%%%%%%%%%%%%%%%%%%%%%%%%%%%%%%%%%%%%%%%%%%%%%%%%%%%%%%%%%%%%%%%%%%%%%%%%%%%%%

\newcommand{\auth}{\large B. Eden\hoch{a}, P.S. Howe\hoch{a},
C. Schubert\hoch{b}, E. Sokatchev\hoch{b} and P.C. West\hoch{a}}

\thispagestyle{empty}

%\begin{document}

\hfill{KCL-MTH-98-58}

\hfill{LAPTH-705/98}

\hfill{hep-th/98mmddnnn}

\hfill{\today}

\vspace{20pt}

\begin{center}
{\Large{\bf Four-point functions in $N=4$ supersymmetric Yang-Mills theory at
two loops}}
\vspace{30pt}

\auth

\vspace{15pt}

\begin{itemize}
\item [$^a$] \kings
\item [$^b$] {\it\small Laboratoire d'Annecy-le-Vieux de Physique
Th{\'e}orique\footnote{URA 1436 associ{\'e}e {\`a} l'Universit{\'e} de Savoie} LAPTH,
Chemin de Bellevue, B.P. 110, F-74941 Annecy-le-Vieux, France}
\end{itemize}

\vspace{60pt}

{\bf Abstract}

\end{center}

Four-point functions of gauge-invariant operators in $D=4, N=4$ supersymmetric
Yang-Mills theory are studied using $N=2$ harmonic superspace perturbation
theory. The results are expressed in terms of differential operators acting on
a scalar two loop integral. The leading singular behaviour is obtained in the
limit that two of the points approach one another. We find logarithmic
singularities which do not cancel out in the sum of all diagrams. It is
confirmed that Green's functions of analytic operators are indeed analytic at
this order in perturbation theory.

{\vfill\leftline{}\vfill

\pagebreak

\section{Introduction}

Supersymmetric Yang-Mills (SYM) theory with maximally extended ($N=4$)
supersymmetry in four dimensions has long been known to have some very
interesting properties. In particular, it is ultra-violet finite and hence
superconformally invariant quantum mechanically \cite{n4}, it admits monopole
solutions which fall into spin one multiplets \cite{os} and it exhibits
Montonen-Olive duality \cite{mo,sen}. More recently, renewed interest in this
theory as a superconformal field theory (SCFT) has been stimulated by the
Maldacena conjecture via which it is related, in a certain limit, to IIB
supergravity (SG) on $AdS_5\times S^5$ \cite{mal}. In this paper we shall study
some $N=4$ SYM four-point correlation functions of gauge-invariant operators at
two loops in perturbation theory. The motivation for this study is threefold:
firstly, from a purely field-theoretic point of view, complete two-loop
four-point calculations in four-dimensional gauge theories are not commonplace;
secondly, it is of interest to see if, even qualitatively, there is any sign of
agreement with supergravity calculations \cite{hong,dzf}, and, thirdly, it
provides a check in perturbation theory of the assumption of references
\cite{hw} that correlation functions of analytic operators are indeed analytic
in the quantum theory.

The main results of the paper concern the evaluation of four-point functions of
gauge-invariant bilinears constructed from $N=2$ hypermultiplets in the adjoint
representation of the gauge group $SU(N_c)$. (We recall that the $N=4$ SYM
multiplet splits into the $N=2$ SYM multiplet and a hypermultiplet in the
adjoint representation.) The calculation of two-loop four-point amplitudes is,
by present standards, still a difficult task in any field theory, except for
quantities which depend only on the ultra-violet divergent parts of diagrams,
such as renormalisation group functions. Very few exact result have been
obtained for the finite parts of such amplitudes, the notable exception being
the case of gluon-gluon scattering in $N = 4$ SYM theory for which the two-loop
on-shell amplitude has been calculated in terms of the scalar double box
integrals \cite{beroya}. Our calculation is carried out in the $N=2$ harmonic
superspace formalism \cite{HSS,frules} and we arrive at answers expressed in
terms of a second-order differential operator acting on the standard $x$-space
scalar two-loop integral. From our results we are then able to deduce the
asymptotic behaviour of the four-point functions as two of the points approach
each other. In this limit the behaviour of the leading singularity can be found
explicitly in terms of elementary functions. We find that it has the form
$(x^2)^{-1} \ln x^2$ where $x$ is the coordinate of the difference between the
two coinciding points. This behaviour, although obtained in perturbation
theory, is in line with the behaviour reported on in references \cite{hong,dzf}
where some progress towards computing four-point functions in supergravity has
been made. We also find purely logarithmic next-to-leading terms.

For the case of three-point functions it is possible to check the Maldacena
conjecture by comparing SG results with perturbation theory for SCFT, because
the higher loop corrections to the leading terms turn out to vanish for the three-supercurrent correlator and vanish at least to first non-trivial order for other correlators
\cite{dzf2,min,hsw}. However, it is more difficult to compare SG results with
perturbative SCFT for higher point functions. One technical difficulty that one
encounters  is that the easiest functions to compute in SCFT are the leading
scalar terms of the four-point function of four supercurrent operators,
whereas, on the SG side, it is easiest to compute amplitudes involving the
axion and dilaton fields. In section 2 we shall show how one may compute the
leading term of the $N=4$ four-point function from the three $N=2$
hypermultiplet four-point functions that are considered in perturbation theory
in section 4. On the grounds that there are no nilpotent superinvariants of
four points \cite{hw} this means that we can, in principle, determine the
entire $N=4$ four-point correlation function from the three $N=2$
hypermultiplet correlation functions. Furthermore, we can also compute the
four-point correlation functions of $N=2$ operators constructed as bilinears in
the $N=2$ SYM field-strength tensor superfield. The leading terms of the
non-vanishing amplitudes of this type are related to the hypermultiplet
correlation functions due to their common origin in $N=4$. In section 3 we then
use a superinvariant argument to show how the $N=2$ YM correlators can be
constructed in their entirety from the leading terms. It is the highest
components of these correlation functions in a $\theta$-expansion which
correspond to the SG amplitudes  currently being investigated.

In the $N=4$ harmonic superspace approach to SYM one can construct a family of
gauge-invariant operators which are described by single-component analytic
superfields. These fields depend on only half the number of odd coordinates of
standard $N=4$ superspace and, in addition, depend analytically on the
coordinates of an auxiliary bosonic space, the coset space $S(U(2)\times
U(2))\backslash SU(4)$, which is a compact complex manifold. Moreover, this
family of operators is in one-to-one correspondence with the KK states of IIB
supergravity compactified on $AdS_5\times S^5$ \cite{af}. In references
\cite{hw} it has been argued that one might hope to get further constraints on
correlation functions of these operators by using superconformal invariance and
the assumption that analyticity is maintained in the quantum theory. However,
this assumption is difficult to check directly in the $N=4$ formalism because
it is intrinsically on-shell. In a recent paper \cite{hsw} analyticity was
checked for certain three-point functions using $N=2$ harmonic superspace
(which is an off-shell formalism). The four-point functions computed in the
current paper also preserve analyticity thereby lending further support to the
$N=4$ analyticity postulate.

The organisation of the paper is as follows: in the next section we shall show
how the leading, scalar term of the $N=4$ four-point correlation function can
be determined from three $N=2$ four-point functions; following this, we show that
knowledge of these functions is also sufficient to determine the $N=2$
four-point function with two $W^2$ operators and two $\bar W^2$ operators, $W$
being the chiral $N=2$ Yang-Mills field-strength multiplet, and we also show
that the leading scalar term of this correlation function can be used, in
principle, to determine it completely. In section 4 we present the ${\cal
N}=2$ harmonic superspace calculations of the two hypermultiplet correlation
functions at two loops in some detail. We then discuss the asymptotic behaviour
of the integrals that occur in these computations in order to find out the
leading singularities that arise when two of the points approach each other.
The paper ends with some further comments. The appendix collects
some known results on massless one-loop and two-loop integrals
which we have found useful.

\section{$N=4$ in terms of $N=2$}

In this section we show how one can compute the leading scalar term of the
$N=4$ four-point function of four supercurrents from the leading scalar terms
of three $N=2$ hypermultiplet four-point functions.  We recall that in standard
$N=4$ superspace the $N=4$ field strength superfield $W^{IJ}=-W^{JI},\
I,J=1\ldots 4$ transforms under the real six-dimensional representation of the
internal symmetry group $SU(4)$ (where the $I$ index labels the fundamental
representation), i.e. it is self-dual, $\bar W_{IJ}={1\over2}\epsilon_{IJKL}
W^{KL}$. This superfield satisfies the (on-shell) constraint

\begin{equation}\label{N4con}
 D_{\alpha}^IW^{JK}=D_{\alpha}^{[I}W^{JK]}
\end{equation}

where $D_{\alpha}^I$ is the superspace covariant derivative. Strictly speaking
this constraint holds only for an Abelian gauge theory and in the non-Abelian
case a connection needs to be included. However, the constraints satisfied by
the gauge-invariant bilinears that we shall consider are in fact the same in
the Abelian and non-Abelian cases.

In order to discuss the energy-momentum tensor it is convenient to use an
$SO(6)$ formalism. The field-strength itself can be written as a vector of
$SO(6)$, $W^A,\ A=1,\ldots 6$,

\be
W^A={1\over2}(\sigma^A)_{JK}W^{JK}
\ee

The sigma-matrices have the following components

\be
\begin{array}{lclc}
(\sigma^a)_{bc}&= 0 & (\sigma^a)_{b4}&=\delta^a_b \\ (\sigma^{\bar
a})_{bc}&=\epsilon_{abc} & (\sigma^{\bar a})_{b4}&= 0
\end{array}
\ee

where the small Latin indices run from 1 to 3, and where the $SO(6)$ and
$SU(4)$ indices split as $A=(a,\bar a)$ (in a complex basis) and $I=(a,4)$. An
upper (lower) $(\bar a)$ index is equivalent to a lower (upper) $a$ index and
vice versa. The sigma-matrices are self-dual,

\be
(\bar\sigma^A)^{IJ}=\frac{1}{2}\epsilon^{IJKL}(\sigma^A)_{KL}
\ee

In terms of the $SU(3)$ indices $a,\bar a$ one has the decompositions,

\be
W^A\rightarrow (W^a, W^{\bar a}=\bar W_a)
\ee

and

\be
W^{IJ}\rightarrow \left(W^{ab}=\epsilon^{abc}\bar W_c,\  W^{a4}=W^a\right)
\ee

The leading component of $W^a$ in an expansion in the fourth $\theta$ can be
identified with the $N=3$ SYM field-strength tensor. Decomposing once more to
$N=2$ one finds that $W^a$ splits into the $N=2$ field-strength $W$ and the
$N=2$ hypermultiplet $\phi^i$,

\be
W^a\rightarrow (W^i\equiv \phi^i, W^3\equiv W)
\ee

{}From (\ref{N4con}) it is easy to see that these superfields (evaluated with
both the third and fourth $\theta$ variables set equal to zero) do indeed obey
the required constraints, that is, $W$ is chiral and

\bea
D_{\alpha}^{(i}\phi^{j)}&=&0\\ \nonumber
D_{\alpha}^{(i}\bar\phi^{j)}&=&0
\eea

In addition, at the linearised level, the SYM field-strength $W$ also satisfies
the equation of motion, $D_{\alpha}^i D^{\alpha j} W=0$, and this also follows
from (\ref{N4con}). The $N=4$ supercurrent is given by \footnote{Here, and
throughout the paper, bilinear expressions such as $WW$, $\phi\phi$ etc. are
understood to include a trace over the Yang-Mills indices.}

\be
T^{AB}=W^A W^B -{1\over6}\delta^{AB}W^C W^C
\ee

and the four-point function we are going to consider is

\be
G^{(N=4)}=<T^{A_1B_1}T^{A_2B_2}T^{A_3B_3}T^{A_4B_4}>
\ee

where the numerical subscripts indicate the point concerned. This function can
be expressed in terms of $SO(6)$ invariant tensors multiplied by scalar factors
which are functions of the coordinates. The only $SO(6)$ invariant tensor that
can arise is $\delta$, and there are two modes of hooking the indices up in
$G^{(N=4)}$ each of which can occur in three combinations making six
independent amplitudes in all. Thus we have

\bea
G^{(N=4)}&=&a_1(\delta_{12})^2(\delta_{34})^2 +
a_2(\delta_{13})^2(\delta_{24})^2
 +a_3(\delta_{14})^2(\delta_{23})^2\\ \nonumber
 &\phantom{=}&
 +b_1\delta_{13}\delta_{14}\delta_{23}\delta_{24}+
 b_2\delta_{12}\delta_{14}\delta_{32}\delta_{34}
 +b_3\delta_{12}\delta_{13}\delta_{42}\delta_{43}
\label{ab}
\eea

where, for example,

\be
(\delta_{12})^2(\delta_{34})^2=\delta_{\{A_1B_1\}}^{A_2B_2}
\delta_{\{A_3B_3\}}^{A_4B_4}
\ee

and

\be
\delta_{13}\delta_{14}\delta_{23}\delta_{24}=\delta_{\{A_1B_1\}}^{\{A_3\{B_4}
\delta_{\{A_2B_2\}}^{A_4\}B_3\}}
\ee

and where the brackets denote tracefree symmetrisation at each point.

In $N=3$ the $N=4$ supercurrent splits into two multiplets, the $N=3$
supercurrent $T^{a\bar b}$, and a second multiplet $T^{ab}$ which contains
amongst other components the fourth supersymmetry current. The $N=3$
supercurrent transforms under the real eight-dimensional representation of
$SU(3)$ while the second multiplet transforms according to the complex
six-dimensional representation. The $N=4$ four-point function decomposes into
several $N=3$ four-point functions. Amongst them we find three which, when
further decomposed under $N=2$, will suffice to determine all the $a$ and $b$
functions in (\ref{ab}):

\bea
G_1^{(N=3)}&=&<T^{a_1b_1}T^{\bar a_2\bar b_2}T^{a_3b_3}T^{\bar a_4\bar b_4}>\\
G_2^{(N=3)}&=&<T^{a_1b_1}T^{\bar a_2\bar b_2}T^{a_3\bar b_3}T^{a_4\bar b_4}>\\
G_3^{(N=3)}&=&<T^{a_1\bar b_1}T^{a_2\bar b_2}T^{a_3\bar b_3}T^{a_4\bar b_4}>
\eea

In the $N=2$ decomposition of $T^{ab}$ we find

\be
T^{ij}=\phi^i\phi^j
\ee

while $T^{33}=W^2$. On the other hand, the $N=3$ supercurrent
contains another $N=2$ hypermultiplet operator, also in a triplet of $SU(2)$.
This is obtained from $T^{a\bar b}$ by restricting the indices to run from 1 to
2 and then removing the trace over the $i,j$ indices. In this way we can
construct

\be
\hat T^i{}_j=T^i{}_j-{1\over2}\delta^i_j
T^k{}_k=\phi^i\bar\phi_j-{1\over2}\delta^i_j\phi^k\bar\phi_k
\ee

The $N=2$ harmonic superspace hypermultiplet $q^+$ is related to the $N=2$
superfield $\phi^i$ by

\be\label{trivsol}
q^+= u^+_i\phi^i
\ee

on-shell. Its conjugate $\widetilde q^+$ is given by

\be
\widetilde q^+= u^{+i}\bar\phi_i
\ee

where the details of $N=2$ harmonic superspace are reviewed in section 4.

Restricting the indices on the three $N=3$ four-point functions $G_1^{(N=3)}$ and
$G_2^{(N=3)}$ to run from 1 to 2, removing the $N=2$ traces at each point where
necessary and multiplying the resulting functions by $u^+_i u^+_j$ at each
point we find three hypermultiplet four-point functions. The leading terms of
these correlation functions are given in terms of the $a$ and $b$ functions
introduced in equation (\ref{ab}) and it is a straightforward computation to show
that, in the notation of section 4,

\bea
G_1^{(N=2)}&=&<\widetilde q^+\widetilde q^+|q^+q^+|\widetilde q^+
\widetilde q^+ |q^+q^+>\\ \nonumber
&=&(12)^2(34)^2 a_1 + (14)^2(23)^2 a_3 + (12)(23)(34)(41) b_2
\eea

while

\bea
G_2^{(N=2)}&=&<\widetilde q^+\widetilde q^+|q^+q^+|q^+\widetilde q^+|q^+\widetilde
q^+> \\ \nonumber &=&{1\over4}\left((12)^2(34)^2(-2a_1 - b_3) + (14)^2(23)^2
b_1+ (12)(23)(34)(41) (b_3-b_1-b_2)\right)
\eea

and

\bea
G_3^{(N=2)}&=&<q^+\widetilde q^+|q^+\widetilde q^+|q^+\widetilde q^+|q^+\widetilde
q^+> \\ \nonumber &=&{1\over8}\Big((12)^2(34)^2(2a_1+2a_2 +b_3) + (14)^2(23)^2
(2a_2 +2a_3 +b_1)+ \\ \nonumber
&\phantom{=}& + (12)(23)(34)(41) (b_2-b_1-b_3-4a_2)\Big)
\eea

where, for example,

\begin{equation}
(12)= u_1^{+i}u_2^{+j}\epsilon_{ij}
\end{equation}

In section 4 we shall see that each of these four-point functions can be
written in terms of three functions of the coordinates, $A_1,A_2,A_3$, and
hence all of the $a$ and $b$ functions that appear in the $N=4$ four-point
function can be determined by computing the above $N=2$ correlators.

We now consider the $N=2$ correlation functions involving the gauge-invariant
operators $W^2$ and $\bar W^2$ constructed from the $N=2$ SYM field-strength.
The possible independent four-point functions of this type are

\begin{eqnarray}
G_4^{(N=2)}&=&<W^2\bar W^2 W^2\bar W^2> \\ \nonumber
G_5^{(N=2)}&=&<W^2 W^2 W^2\bar W^2>
\\ \nonumber
G_6^{(N=2)}&=&<W^2 W^2 W^2 W^2>
\end{eqnarray}

The leading terms of $G_5^{(N=2)}$ and $G_6^{(N=2)}$ vanish as one can easily
see by examining the leading terms of the $N=3$ correlation functions from
which they can be derived. In fact, it is to be expected that these correlation
functions should vanish at all orders in $\theta$. This can be argued from an
$N=4$ point of view from the absence of nilpotent $N=4$ superinvariants, or
directly in $N=2$. For example, using just $N=2$ superconformal invariance, it
is possible to show that all correlation functions of gauge-invariant powers of
$W$ vanish.

We remark that the $N=2$ four-point function $G_4^{(N=2)}=<W^2\bar W^2W^2\bar
W^2>$ is also obtainable from $G_2^{(N=3)}$. In terms of the $a$ and $b$
functions it is given by

\begin{equation}
<W^2\bar W^2W^2\bar W^2>=a_1 + a_3 + b_2
\end{equation}

On-shell the $(\theta)^4$ component of $W^2$ is
$F_{\alpha\beta}F^{\alpha\beta}$ where $F_{\alpha\beta}$ is the self-dual
space-time Yang-Mills field-strength tensor, so that the top $((\theta)^{16})$
component of this correlation function is directly related to the dilaton and
axion amplitudes in supergravity. Clearly any four-point function of the
operators $F_{\mu\nu}F^{\mu\nu}$ and ${1\over
2}\epsilon^{\mu\nu\rho\sigma}F_{\mu\nu}F_{\rho\sigma}$ can be obtained from the
above.

\section{The $N=2$ chiral-antichiral four-point function}

In this section we show that the complete $N=2$ four-point function
$G_4^{(N=2)}$ is determined by its leading term using only superconformal
invariance. The coordinates of $N=2$ superspace are
$(x^{\alpha\dot\alpha},\theta^{\alpha}_i,\bar\theta^{\dot\alpha i})$ and an
infinitesimal superconformal transformation in this space is given by a
superconformal Killing vector field
$V=F^{\alpha\dot\alpha}\partial_{\alpha\dot\alpha}+\varphi^{\alpha}_i
D_{\alpha}^i-\bar \varphi^{\dot \alpha i}\bar D_{\dot \alpha i}$, where
$D_{\alpha}^i$ is the usual flat space supercovariant derivative. By
definition, $V$ satisfies the equation

\begin{equation}
[D_{\alpha}^i,V]\cong D_{\alpha}^i
\end{equation}

The chiral superfield $W^2$ transforms as

\begin{equation}
\delta W^2=V W^2 + \Delta W^2
\end{equation}

where $\Delta=\partial_{\alpha\dot\alpha}F^{\alpha\dot\alpha}-D_{\alpha
}^i\varphi^{\alpha}_i$.

Since $G_4^{(N=2)}$ is chiral at points 1 and 3 and anti-chiral at points 2 and
4 it depends only on the chiral or anti-chiral coordinates at these points.
These are given by

\begin{equation}
  \begin{array}{lclcll}
    X^{\alpha\dot\alpha}&=&x^{\alpha\dot\alpha}+
    2{i}\theta^{\alpha}_i\bar\theta^{\dot\alpha i}
    &;&\theta^{\alpha}_i &{\rm chiral} \\
    \bar X^{\alpha\dot\alpha}&=&x^{\alpha\dot\alpha}-2{i}\theta^{\alpha}_i
    \bar\theta^{\dot\alpha i}
    &;&\bar\theta^{\dot\alpha i}&{\rm anti-chiral}
  \end{array}
\end{equation}

At lowest order in Grassmann variables translational invariance in $x$ implies
that $G_4^{(N=2)}$ depends only on the difference variables $x_r- x_s,\
r,s=1\ldots 4$, of which three are independent. Combining this with
$Q$-supersymmetry one finds that $G_3^{(N=2)}$ depends only on the
$Q$-supersymmetric extensions of these differences, which will be denoted
$y_{rs}$, as well as the differences $\theta_{13}=\theta_1-\theta_3$ and
$\bar\theta_{24}=\bar\theta_2-\bar\theta_4$. The supersymmetric difference
variables joining one chiral point $(r)$ with one anti-chiral point $(s)$ have
the following form:

\begin{equation}
y_{rs}=X_r-\bar X_s -4i\theta_r\bar\theta _s
\end{equation}

When the chirality of the two points is the same one has

\begin{eqnarray}
 y_{13}&=&X_1-X_3 -2{i}\theta_{13}(\bar\theta_2 +\bar\theta_4)\\
 \nonumber
 y_{24}&=&\bar X_2-\bar X_4 +2{i}(\theta_1 +\theta_3)\bar\theta_{24}
\end{eqnarray}

It is easy to find a free solution of the Ward Identity for $G_4^{(N=2)}$; it is
given by

\begin{equation}
G_4^{o(N=2)}={1\over y_{14}^2 y_{23}^2}
\end{equation}

A general solution can be written in terms of this free solution in the form

\begin{equation}
G_4^{(N=2)}=G_4^{o(N=2)} \times F
\end{equation}

where $F$ is a function of superinvariants. At the lowest order, $F$ is a
function of two conformal invariants which may be taken to be

\begin{equation}
 S\equiv{x_{12}^2x_{34}^2\over x_{14}^2x_{23}^2};\qquad
 T\equiv{x_{13}^2x_{24}^2\over x_{14}^2x_{23}^2}
\end{equation}

The strategy is now to show that these two conformal invariants can be extended
to superconformal invariants and furthermore that there are no further
superconformal invariants. Any new superinvariant would have to vanish at
lowest order and would thus be nilpotent.

{}From the above discussion it is clear that $F$ can only depend on the $y_{rs}$
and the odd differences $\theta_{13}$ and $\bar\theta_{24}$. Furthermore, it
has dilation weight and $R$-weight zero, the latter implying that it depends on
the odd coordinates in the combination $\theta_{13}\bar\theta_{24}$. This takes
all of the symmetries into account except for $S$-supersymmetry and conformal
symmetry. However, conformal transformations appear in the commutator of two
$S$-supersymmetry transformations and so it is sufficient to check the latter.
The $S$-supersymmetry transformations of the chiral-antichiral variables are
rather simple:

\begin{equation}
 \delta y_{rs}^{\alpha\dot\alpha}=
 \theta_{r i}^{\alpha}\eta_{\beta}^iy_{rs}^{\beta\dot\alpha}
\end{equation}

The transformations of the (anti-)chiral-chiral variables are slightly more
complicated and it is convenient to introduce new variables as follows:

\begin{eqnarray}
\hat y_{13}&=& y_{12}+y_{23}\\ \nonumber
\hat y_{24}&=& -y_{12}+y_{14}
\end{eqnarray}

Under $S$-supersymmetry these transform as follows:

\begin{eqnarray}
\delta\hat y_{13}^{\alpha\dot\alpha}&=&
\theta_{3\hspace{1 pt} i}^{\alpha}\eta_{\beta}^i\hat y_{13}^{\beta\dot\alpha}
+\theta_{13\hspace{1 pt} i}^{\alpha}\eta_{\beta}^i y_{12}^{\beta\dot\alpha}\nonumber\\
\delta\hat y_{24}^{\alpha\dot\alpha}&=&
\theta_{1\hspace{1 pt} i}^{\alpha}\eta_{\beta}^i\hat y_{24}^{\beta\dot\alpha}
\end{eqnarray}

The transformations of the odd variables are

\begin{eqnarray}
\delta\theta^{\alpha}_i&=&\theta^{\alpha}_j\eta_{\beta}^j\theta^{\beta}_i\\
\nonumber
\delta\bar\theta^{\dot\alpha i}&=&-{i\over4}\eta_{\alpha}^i
\bar X^{\alpha\dot\alpha}
\end{eqnarray}

Using these transformations it is easy to extend $S$ to a superinvariant $\hat S$.  It is

\begin{equation}
 \hat S={y_{12}^2y_{34}^2\over y_{14}^2y_{23}^2}
\end{equation}

However, the extension of the second invariant is slightly more complicated
because it involves the (anti-)chiral-chiral differences. A straightforward
computation shows that the required superinvariant is

\begin{equation}
\hat T={1\over y_{14}^2y_{23}^2}\left(\hat y_{13}^2\hat y_{24}^2
-16i\theta_{13}\cdot\bar\theta_{24}\cdot(\hat y_{13}y_{12}\hat y_{24})-
16y_{12}^2(\theta_{13}\cdot\bar\theta_{24})^2\right)
\end{equation}

Now consider the possibility of a nilpotent superinvariant. Under an
$S$-supersymmetry transformation the leading term in the variation arises from
the $\theta$-independent term in the variation of $\bar\theta$. On the
assumption that the $x$-differences are invertible, it follows immediately that
there can be no such invariants. Thus, the general form of the four-point
function $G_4^{(N=2)}$ is

\begin{equation}
<W^2\bar W^2W^2\bar W^2>={1\over y_{14}^2y_{23}^2}F(\hat S,\hat T)
\end{equation}

and so is calculable from the leading term as claimed.

\section{Computation of four-point $N=2$ correlation functions}

\subsection{$N =2$ harmonic superspace and Feynman rules}

\subsubsection{}

The $N =4$ SYM multiplet reduces to an $N =2$ SYM multiplet and a
hypermultiplet. The latter are best described in $N =2$ harmonic superspace
\cite{HSS}. In addition to the usual bosonic ($x^{\alpha\dot\alpha}$) and
fermionic ($\theta^{\alpha}_i, \bar\theta^{\dot\alpha i}$) coordinates, it
involves $SU(2)$ harmonic ones:

\begin{equation}\label{harco}
SU(2)\ni u=(u_i^+,u_i^-)\;: \qquad u^-_i =\overline{u^{+i}}\;, \quad
u^{+i}u^-_i
= 1\ .
\end{equation}

They parametrise the coset space $SU(2)/U(1)\sim S^2$ in the following sense:
the index $i$ transforms under the left $SU(2)$ and the index (``charge") $\pm$
under the right $U(1)$; further, all functions of $u^\pm$ are homogeneous under
the action of the right $U(1)$ group. The harmonic functions $f(u)$ are defined
by their harmonic expansion (in general, infinite) on $S^2$.

The main advantage of harmonic superspace is that it allows us to define
Grassmann-analytic (G-analytic) superfields satisfying the constraints

\be\label{Gan}
D^+_\alpha \Phi(x,\theta,u) = \bar D^+_{\dot\alpha} \Phi(x,\theta,u) = 0\;.
\ee

Here

\be
D^+_{\alpha,\dot\alpha} = u^+_i D^i_{\alpha,\dot\alpha}
\ee

are covariant $U(1)$ harmonic projections of the spinor derivatives. The
G-analyticity condition (\ref{Gan}) is integrable, since by projecting the
spinor derivative algebra

\be
\{D^i_\alpha,D^j_\beta\} = \{\bar D_{\dot\alpha i},\bar D_{\dot\beta j}\} = 0,
\quad
\{D^i_\alpha,\bar D_{\dot\beta j}\} = -2i
\delta^i_j \partial_{\alpha\dot\beta}
\ee

one finds

\be
\{D^+_\alpha,D^+_\beta\} = \{\bar D^+_{\dot\alpha},\bar D^+_{\dot\beta}\} =
\{D^+_\alpha,\bar D^+_{\dot\beta}\} = 0\;.
\ee

Moreover, the G-analyticity condition (\ref{Gan}) can be solved explicitly. To
this end one introduces a new, analytic basis in harmonic superspace:

\be\label{anbas}
x^{\alpha\dot\alpha}_A = x^{\alpha\dot\alpha} - 2i\theta^{\alpha (i}
\bar\theta^{\dot\alpha j)} u^+_i u^-_j\;, \quad \theta^\pm_{\alpha,\dot\alpha} =
u^\pm_i\theta^i_{\alpha,\dot\alpha}\;, \quad u^\pm_i \;.
\ee

In this basis the constraints (\ref{Gan}) just imply

\be\label{Gansupf}
\Phi^q = \Phi^q(x_A,\theta^+,\bar\theta^+,u^\pm)\;,
\ee

i.e. the solution is a Grassmann-analytic function of $\theta^+,\bar\theta^+$
(in the sense that it does not depend on the complex conjugates
$\theta^-,\bar\theta^-$).

We emphasise that the superfield (\ref{Gansupf}) is a non-trivial harmonic
function carrying an external $U(1)$ charge $q=0,\pm 1,\pm 2,\ldots$. Thus,
$\Phi^q$ has an infinite harmonic expansion on the sphere $S^2$. Most of the
terms in this expansion turn out to be auxiliary (in the case of the
hypermultiplet) or pure gauge (in the SYM case) degrees of freedom. In order to
obtain an ordinary superfield with a finite number of components one has to
restrict the harmonic dependence. This is done with the help of the harmonic
derivative (covariant derivative on $S^2$)

\be
D^{++} = u^{+i}{\partial\over\partial u^{-i}}\;,
\ee

or, in the analytic basis (\ref{anbas}),

\be\label{GanD}
D^{++} = u^{+i}{\partial\over\partial u^{-i}} -2i\theta^{+\alpha}
\bar\theta^{+\dot\alpha}
{\partial\over\partial
x^{\alpha\dot\alpha}_A}\;.
\ee

Thus, for instance, one can choose the G-analytic superfield
$q^+(x_A,\theta^+,\bar\theta^+,u^\pm)$ carrying $U(1)$ charge $+1$ and impose
the harmonic analyticity condition

\be\label{emhyp}
D^{++}q^+=0\;.
\ee

Due to the presence of a space-time derivative in the analytic form
(\ref{GanD}) of $D^{++}$, equation (\ref{emhyp}) not only ``shortens'' the
harmonic superfield $q^+$, but also puts the remaining components on shell. In
fact, (\ref{emhyp}) is the equation of motion for the $N =2$ hypermultiplet. We
remark that eq. (\ref{emhyp}) is compatible with the G-analyticity conditions
(\ref{Gan}) owing to the obvious commutation relations

\be
[D^{++}, D^+_{\alpha,\dot\alpha}] = 0\;.
\ee

Note also that in the old basis $(x,\theta_i,\bar\theta^i,u)$ the harmonic
analyticity condition has the trivial solution given in (\ref{trivsol});
however, in this basis the G-analyticity condition (\ref{Gan}) on $q^+$ becomes
a non-trivial condition which in fact follows from the $N=4$ SYM on-shell
constraints (\ref{N4con}).

\subsubsection{}

A remarkable feature of harmonic superspace is that it allows us to have an
off-shell version of the hypermultiplet. To this end it is sufficient to relax
the on-shell condition (\ref{emhyp}) and subsequently obtain it as a
variational equation from the action

\be\label{HMact}
S_{HM} = \int d^4x_A du d^4\theta^+\; \widetilde q^+ D^{++} q^+\;.
\ee

Here the integral is over the G-analytic superspace: $du$ means the standard
invariant measure on $S^2$ and $d^4\theta^+\equiv (D^-)^4$. The special
conjugation $\widetilde q^+$ combines complex conjugation and the antipodal map
on $S^2$ \cite{HSS}. Its essential property is that it leaves the G-analytic
superspace invariant (unlike simple complex conjugation). The reality of the
action is established with the help of the conjugation rules $\widetilde
{\widetilde q^+} = - q^+$ and $\widetilde D^{++} = D^{++}$. Note that the
$U(1)$ charge of the Lagrangian in (\ref{HMact}) is $+4$, which exactly matches
that of the measure (otherwise the $SU(2)$ invariant harmonic integral would
vanish).

The other ingredient of the $N=4$ theory is the $N=2$ SYM multiplet. It is
described by a real G-analytic superfield
$V^{++}(x_A,\theta^+,\bar\theta^+,u^\pm) = \widetilde V^{++}$ of $U(1)$ charge
$+2$ subject to the following gauge transformation:

\be\label{gtrv}
\delta V^{++} = -D^{++}\Lambda +ig[\Lambda,V^{++}]
\ee

where $\Lambda(x_A,\theta^+,\bar\theta^+,u^\pm)$ is a G-analytic gauge
parameter. It can be shown \cite{HSS} that what remains from the superfield
$V^{++}$ in the Wess-Zumino gauge is just the (finite-component) $N=2$ SYM
multiplet. Under a gauge transformation (\ref{gtrv}), the matter
(hypermultiplet) superfields transforms in the standard way

\be
\delta q^+ = i\Lambda q^+\; .
\ee

Thus $V^{++}$ has the interpretation of the gauge connection for the harmonic
covariant derivative ${\cal D}^{++} = D^{++} + igV^{++}$. All this suggests the
standard minimal SYM-to-matter coupling which consists in covariantising the
derivative in (\ref{HMact}):

\be\label{HMSYMact}
S_{HM/SYM} = \int d^4x_A du d^4\theta^+\; \left[\widetilde q^+_a (\delta_{ab}
D^{++} + {g\over 2}f_{abc}V^{++}_c) q^+_b\right]\;.
\ee

In order to reproduce the $N=4$ theory we have chosen the matter $q^+ = q^+_a
t_a$ in the adjoint representation of the gauge group $G$ with structure
constant $f_{abc}$
\footnote{We use the definitions $[t_a,t_b] = if_{abc}t_c$, $\mbox{tr}(t_at_b) =
C(Adj)\delta_{ab}$, where the quadratic Casimir is normalised so that $C(Adj)
= N_c$ for $SU(N_c)$.}.

We shall not explain here how to construct the gauge-invariant action for
$V^{++}$ \cite{HSS}. We only present the form of the gauge-fixed kinetic term
in the Fermi-Feynman gauge \cite{frules}:

\be\label{FF}
S_{SYM+GF} = -{1\over 2}\int d^4x_A du d^4\theta^+\; V^{++}_a\Box V^{++}_a\;.
\ee

Of course, the full SYM theory includes gluon vertices of arbitrary order as
well as Faddeev-Popov ghosts, but we shall not need them here (the details can
be found in \cite{frules}).

\subsubsection{}

The off-shell description of both theories above allows us to develop
manifestly $N=2$ supersymmetric Feynman rules. We begin with the gauge
propagator. Since the corresponding kinetic operator in the FF gauge (\ref{FF})
is simply $\Box$, the propagator is given by the Green's function
$1/4i\pi^2x^2_{12}$:

\be\label{invbox}
\Box_1 {1\over 4i\pi^2x^2_{12}} =  \delta(x_1-x_2)
\ee

combined with the appropriate Grassmann and harmonic delta functions:

\vskip 0.5 in
\begin{center}
\begin{picture}(30000,4000)(0,-2000)
\put(0,2000){\scriptsize{1a}}
\drawline\gluon[\E\REG](1500,2000)[6]
\put(8430,2000){\scriptsize{2b}}
\put(15000,2000){$\langle V^{++}_a(1)V^{++}_b(2)\rangle =$}
\put(1500,-1500){Figure 1}
\end{picture}
\end{center}

\be\label{Vprop}
 = {i\over 4\pi^2} \delta_{ab} (D^+_1)^4
\left({\delta_{12}\over
x^2_{12}}\right)
\delta^{(-2,2)}(u_1,u_2) =  {i\over 4\pi^2} \delta_{ab} (D^+_2)^4
\left({\delta_{12}\over
x^2_{12}}\right)
\delta^{(2,-2)}(u_1,u_2)\;,
\ee

where $\delta_{12}$ is shorthand for the Grassmann delta function
$\delta^8(\theta_1-\theta_2)$ and $\delta^{(2,-2)}(u_1,u_2)$ is a harmonic
delta function carrying $U(1)$ charges $+2$ and $-2$ with respect to its first
and second arguments. Note that the propagator is written down in the usual
basis in superspace and not in the analytic one, so $x_{1,2}$ appearing in
(\ref{Vprop}) are the ordinary space-time variables and not the shifted $x_A$
from (\ref{anbas}). The G-analyticity of the propagator is assured by the
projector

\be
(D^+)^4 = {1\over 16} D^{+\alpha}D^+_\alpha \bar D^+_{\dot\alpha} \bar
D^{+\dot\alpha}\;.
\ee

The two forms given in (\ref{Vprop}) are equivalent owing to the presence of
Grassmann and harmonic delta functions \footnote{In certain cases the harmonic
delta function needs to be regularised in order to avoid coincident harmonic
singularities. In such cases one should use an equivalent form of the
propagator (\ref{Vprop}) in which the analyticity with respect to both
arguments is manifest \cite{Ky}.}.

The matter propagator is somewhat more involved. The kinetic operator for the
hypermultiplet is a harmonic derivative, so one should expect a harmonic
distribution to appear in the propagator. Such distributions are simply given
by the inverse powers of the $SU(2)$ invariant combination $u^{+i}_1u^+_{2i}
\equiv (12)$ which vanishes when $u_1 = u_2$. One can prove the relations
\cite{frules}:

\be\label{hdistr}
D^{++}_1 {1\over (12)^n} = {1\over (n-1)!}
(D^{--}_1)^{n-1}\delta^{(n,-n)}(u_1,u_2)
\ee

which are in a way the $S^2$ analogues of eq. (\ref{invbox}). Here $D^{--} =
\overline{D^{++}}$ is the other covariant derivative
on $S^2$. So, the $q^+$ propagator is then given by

\vskip 0.5 in
\begin{center}
\begin{picture}(30000,4000)(0,-2000)
\put(0,2000){\scriptsize{1a}}
\drawline\fermion[\E\REG](1500,2000)[6430]
\drawarrow[\E\ATTIP](\pmidx,\pmidy)
\put(8430,2000){\scriptsize{2b}}
\put(15000,2000){$\langle \widetilde q^+_a(1) q^+_b(2)\rangle =$}
\put(1500,-1500){Figure 2}
\end{picture}
\end{center}

\be\label{qprop}
  = {i\over 4\pi^2} \delta_{ab}
{(D^+_1)^4(D^+_2)^4\over (12)^3}
\left({\delta_{12}\over x^2_{12}}\right)  \equiv \Pi_{12} \delta_{ab} \;.
\ee

This time we need the presence of two G-analyticity projectors
$(D^+_1)^4(D^+_2)^4$ because we do not have a harmonic delta function any more.
In order to show that (\ref{qprop}) is indeed the Green's function for the
operator $D^{++}$, one uses (\ref{hdistr}) and the identity

\be
-{1\over 2} (D^+)^4 (D^{--})^2 \Phi = \Box \Phi
\ee

on any G-analytic superfield $\Phi$.

Finally, the only vertex relevant to our two-loop calculation can be read off
from the
coupling term in (\ref{HMSYMact}):

\vskip 1 in
\begin{center}
\begin{picture}(30000,4000)(0,-4000)
\drawline\fermion[\E\REG](0,0)[3215] 
\drawarrow[\E\ATTIP](\pmidx,\pmidy) 
\drawline\gluon[\N\CENTRAL](\fermionbackx,\fermionbacky)[3] 
\put(-1000,\fermionbacky){\scriptsize{a}} 
\put(7000,\fermionbacky){\scriptsize{b}} 
\put(3000,4500){\scriptsize{c}} \put(2815,-1500){\scriptsize{1}} 
\drawline\fermion[\E\REG](\fermionbackx,\fermionbacky)[3215] 
\drawarrow[\E\ATTIP](\pmidx,\pmidy) \put(11000,2000){${g\over 2} 
f_{abc}\int d^4x_{A,1} du_1 d^4\theta_{1}^+$} 
\put(1500,-3500){Figure 3}\label{vertex} 
\end{picture}
\end{center}
It involves an integral over the G-analytic superspace. 

Note also the following useful relations. The full superspace 
Grassmann measure is related to the G-analytic one by 

\be\label{fullmeas} d^8\theta = d^4\theta^+ (D^+)^4 = 
(D^-)^4(D^+)^4\;. \ee 

The Grassmann delta function $\delta^8(\theta_1-\theta_2) \equiv 
\delta_{12}$ is defined as usual, 

\be
\int d^8\theta\; \delta^8(\theta) =\left. 
(D^-)^4(D^+)^4\delta^8(\theta)\right\vert_{\theta=0} = 1\;, \ee 

from which it is easy to derive 

\be
\left.(D^+_1)^4(D^+_2)^4\delta_{12}\right\vert_{\theta=0} = 
(12)^4\;. \ee 

Using this relation as a starting point we find others, e.g.: 

\bea 
\left.(D^+_3)^2(D^+_1)^4(D^+_2)^4\delta_{12}\right\vert_{\theta=0} 
&=& \left.(\bar 
D^+_3)^2(D^+_1)^4(D^+_2)^4\delta_{12}\right\vert_{\theta=0} = 0\;, 
\nonumber\\ \left.D^+_{3\alpha}\bar 
D^+_{3\dot\alpha}(D^+_1)^4(D^+_2)^4\delta_{12}\right\vert_{\theta=0} 
&= & 
-2i(13)(23)(12)^3\partial_{1\alpha\dot\alpha}\;,\label{relations} 
\\ 
\left.(D^+_3)^4(D^+_1)^4(D^+_2)^4\delta_{12}\right\vert_{\theta=0} 
&=& - (13)^2(23)^2(12)^2\Box_1\;, \quad \mbox{etc.} \nonumber \eea 

\subsection{Four-point hypermultiplet correlators}

In what follows we shall apply the above Feynman rules to compute 
four-point correlators of composite gauge invariant operators made 
out of two hypermultiplets. There are two types of such composite 
operators: $q^+q^+$ (and its conjugate $\widetilde q^+\widetilde 
q^+$) and $\widetilde q^+q^+$. The structure of the hypermultiplet 
propagator (\ref{qprop}) suggests that we need equal numbers of 
$q^+$'s and $\widetilde q^+$'s in order to form a closed 
four-point loop. Indeed, for instance, correlators of the type 

\be\label{0corr} \langle q^+q^+\vert q^+q^+\vert q^+q^+\vert 
q^+q^+\rangle \ee 

must vanish in the free case as well as to all orders in 
perturbation theory. The reason is that the only interaction the 
$q^+$'s have is given by the vertex (\ref{vertex}) and it is easy 
to see that there are no possible graphs of this type. The same 
applies to any configuration with unequal numbers of $q^+$'s and 
$\widetilde q^+$'s. So, the non-trivial ones are 

\be\label{1corr} \langle \widetilde q^+\widetilde q^+\vert 
q^+q^+\vert \widetilde q^+\widetilde q^+\vert q^+q^+\rangle\;, \ee 

\be\label{3corr} \langle \widetilde q^+\widetilde q^+\vert 
q^+q^+\vert \widetilde q^+ q^+\vert \widetilde q^+q^+\rangle\;, 
\ee 

\be\label{2corr} \langle \widetilde q^+q^+\vert \widetilde 
q^+q^+\vert \widetilde q^+q^+\vert \widetilde q^+q^+\rangle\;. \ee 

As explained in section 2, the correlators 
(\ref{1corr})-(\ref{2corr}) are sufficient to determine the full 
correlator of four $N=4$ supercurrents. In fact, as we shall see 
later on, it is enough to compute (\ref{1corr}), the other two can 
then be obtained by permutations of the points and symmetrisation. 

The relevant graph topologies for the computation of the 
correlator (\ref{1corr}) are shown in Figure 4: 

\vskip 2 in 
\begin{center}
\begin{picture}(42000,7000)(0,-12000)

\drawline\fermion[\E\REG](0,0)[10000] \global\advance\pmidx by 
-400 \global\Yone=-1500 \put(\pmidx,\Yone){a} 
\global\advance\pmidx by 400 
\drawline\gluon[\N\CENTRAL](\pmidx,\pmidy)[6] 
\global\Xone=\gluonlengthy 
\drawline\fermion[\W\REG](\gluonbackx,\gluonbacky)[5000] 
\drawline\fermion[\E\REG](\gluonbackx,\gluonbacky)[5000] 
\drawline\fermion[\S\REG](\pbackx,\pbacky)[\Xone] 
\drawline\fermion[\N\REG](0,0)[\Xone] 

\drawline\fermion[\E\REG](15000,0)[10000] \global\advance\pmidx by 
-400 \put(\pmidx,\Yone){b} 
\drawline\fermion[\N\REG](\pbackx,\pbacky)[\Xone] 
\drawline\fermion[\W\REG](\pbackx,\pbacky)[10000] 
\drawline\fermion[\S\REG](\pbackx,\pbacky)[\Xone] 
\global\Xtwo=\pmidx \global\Ytwo=\pfronty \startphantom 
\drawline\gluon[\NE\FLIPPED](\pmidx,\pmidy)[3] \stopphantom 
\global\Ythree=\gluonlengthy \global\negate\Ythree 
\global\advance\Ytwo by \Ythree 
\drawline\gluon[\NE\FLIPPED](\Xtwo,\Ytwo)[3] 

\startphantom \drawloop\gluon[\N 5](30000,0) \stopphantom 
\global\Xfive=\loopfrontx \global\negate\Xfive 
\global\advance\Xfive by \loopbackx \global\advance\Xfive by 
-10000 \global\divide\Xfive by 2 
\drawline\fermion[\E\REG](30000,0)[10000] \global\advance\pmidx by 
-400 \put(\pmidx,\Yone){c} 
\drawline\fermion[\N\REG](\pbackx,\pbacky)[\Xone] 
\drawline\fermion[\W\REG](\pbackx,\pbacky)[10000] 
\global\advance\pfrontx by \Xfive \drawloop\gluon[\S 
5](\pfrontx,\pfronty) \drawline\fermion[\N\REG](30000,0)[\Xone] 

\global\Xfour=\Xone \global\multroothalf\Xfour 
\drawline\fermion[\N\REG](5000,-10000)[\Xone] 
\drawline\fermion[\NE\REG](5000,-10000)[\Xfour] 
\drawline\fermion[\NW\REG](\pbackx,\pbacky)[\Xfour] 
\drawline\gluon[\E\REG](\pfrontx,\pfronty)[5] \global\advance\Yone 
by -10000 \global\advance\pmidx by -400 \put(\pmidx,\Yone){d} 
\drawline\fermion[\SE\REG](\pbackx,\pbacky)[\Xfour] 
\drawline\fermion[\N\REG](\pbackx,\pbacky)[\Xone] 
\drawline\fermion[\SW\REG](\pbackx,\pbacky)[\Xfour] 

\drawline\fermion[\NE\REG](35000,-10000)[\Xfour] 
\drawline\fermion[\NW\REG](\pbackx,\pbacky)[\Xfour] 
\drawline\fermion[\SW\REG](\pbackx,\pbacky)[\Xfour] 
\drawline\fermion[\SE\REG](\pbackx,\pbacky)[\Xfour] 
\drawline\gluon[\W\FLIPPED](\pfrontx,\pfronty)[5] 
\global\advance\pmidx by -400 \put(\pmidx,\Yone){e} 
\global\Xsix=\pbackx \global\advance\Xsix by -2500 
\global\Xseven=\pbackx \global\advance\Xseven by -5000 
\global\Ysix=\pbacky \global\advance\Ysix by 900 
\global\Yseven=\pbacky 
\curve(\pbackx,\pbacky,\Xsix,\Ysix,\Xseven,\Yseven) 
\global\advance\Ysix by -1800 
\curve(\pbackx,\pbacky,\Xsix,\Ysix,\Xseven,\Yseven) 

\global\advance\Yone by -1500 \put(15800,\Yone){Figure 4} 
\end{picture}
\end{center}

The graph (c) contains a vanishing two-point insertion (see 
\cite{Ky}). The graphs (d) and (e) are proportional to the trace 
of the structure constant $f_{abb}$ and thus vanish unless the 
gauge group contains a $U(1)$ factor \footnote{The hypermultiplet 
in the $N=4$ multiplet has no electric charge, therefore a $U(1)$ 
gauge factor corresponds to a trivial free sector of the theory. 
Note, nonetheless, that if the graphs (d) and (e) are to be 
considered, they contain divergent $x$-space integrals.}. Thus, we 
only have to deal with the topologies (a) and (b). We shall do the 
calculation in some detail for the case (a), the other one being 
very similar. 

Here is a detailed drawing of the configurations having the 
topology of graph (a): 

\vskip 0.5 in 
\begin{center}
\begin{picture}(38000,10000)(0,-4000)

\startphantom \drawline\gluon[\E\CENTRAL](0,0)[14] \stopphantom 
\global\Xone=\gluonlengthx \global\Xeight=\Xone 
\global\divide\Xeight by 4 \drawline\fermion[\E\REG](0,0)[\Xone] 
\drawarrow[\W\ATTIP](\Xeight,0) \global\multiply\Xeight by 3 
\drawarrow[\W\ATTIP](\Xeight,0) \put(400,200){4} 
\global\advance\pmidx by -400 \global\Ytwo=-1500 
\put(\pmidx,\Ytwo){$I_1$} \global\advance\pmidx by 800 
\put(\pmidx,200){6} \global\advance\pmidx by -400 
\global\advance\pbackx by 400 \put(\pbackx,200){3} 
\drawline\gluon[\N\CENTRAL](\pmidx,\pmidy)[8] 
\global\advance\pmidx by -300 \global\advance\pmidy by 270 
%\drawarrow[\SE\ATTIP](\pmidx,\pmidy)
\global\Yone=\gluonlengthy \drawarrow[\E\ATTIP](\Xeight,\Yone) 
\global\divide\Xeight by 3 \drawarrow[\E\ATTIP](\Xeight,\Yone) 
\global\advance\Yone by 200 \global\advance\pbackx by 400 
\put(\pbackx,\Yone){5} \global\divide\Xone by 2 
\drawline\fermion[\W\REG](\gluonbackx,\gluonbacky)[\Xone] 
\global\advance\pbackx by 400 \put(\pbackx,\Yone){1} 
\drawline\fermion[\E\REG](\gluonbackx,\gluonbacky)[\Xone] 
\global\advance\pbackx by 400 \put(\pbackx,\Yone){2} 
\global\advance\Yone by -200 
\drawline\fermion[\S\REG](\fermionbackx,\pbacky)[\Yone] 
\drawarrow[\N\ATTIP](\pmidx,\pmidy) 
\drawline\fermion[\N\REG](0,0)[\Yone] 
\drawarrow[\S\ATTIP](\pmidx,\pmidy) 

\global\multiply\Xone by 2 
\drawline\fermion[\E\REG](22840,0)[\Xone] 
\drawarrow[\W\ATTIP](\pmidx,\pmidy) 
\drawarrow[\E\ATTIP](\pmidx,\Yone) \put(22900,200){4} 
\global\advance\pmidx by -400 \put(\pmidx,\Ytwo){$I_2$} 
\global\advance\pbackx by 400 \put(\pbackx,200){3} 
\drawline\fermion[\N\REG](\fermionbackx,\pbacky)[\Yone] 
\global\Yeight=\Yone \global\divide\Yeight by 4 
\drawarrow[\S\ATTIP](22840,\Yeight) 
\drawarrow[\N\ATTIP](\pbackx,\Yeight) \global\multiply\Yeight by 3 
\drawarrow[\S\ATTIP](22840,\Yeight) 
\drawarrow[\N\ATTIP](\pbackx,\Yeight) \global\advance\pmidx by 400 
\global\advance\pmidy by 200 \put(\pmidx,\pmidy){5} 
\global\advance\pbackx by 400 \global\advance\pbacky by 200 
\put(\pbackx,\pbacky){2} 
\drawline\fermion[\W\REG](\fermionbackx,\fermionbacky)[\Xone] 
\global\advance\pbackx by 400 \global\advance\pbacky by 200 
\put(\pbackx,\pbacky){1} 
\drawline\fermion[\S\REG](\fermionbackx,\fermionbacky)[\Yone] 
\drawline\gluon[\E\CENTRAL](\pmidx,\pmidy)[14] 
\global\advance\pmidx by 260 \global\advance\pmidy by 200 
%\drawarrow[\SW\ATTIP](\pmidx,\pmidy)
\global\advance\pfrontx by -800 \global\advance\pfronty by 200 
\put(\pfrontx,\pfronty){6} 

\put(15800,-3000){Figure 5} 
\end{picture}
\end{center}

The expression corresponding to the first of them is (up to a 
factor containing the 't Hooft parameter $g^2N_c$): 

\be
I_1 = -\Pi_{14}\Pi_{32}\int d^4x_5 d^4x_6 du_5du_6 d^4\theta^+_5 
d^4\theta^+_6 \Pi_{15}\Pi_{52}\Pi_{36}\Pi_{64} (D^+_6)^4 
\left({\delta_{56}\over x^2_{56}}\right) \delta^{(2,-2)}(u_5,u_6) 
\;. \nonumber \ee 

The technique we shall use to evaluate this graph is similar to 
the usual $D$-algebra method employed in $N=1$ supergraph 
calculations. First, since the propagators $\Pi$ are G-analytic, 
we can use the four spinor derivatives $(D^+_6)^4$ to restore the 
full Grassmann vertex $d^8\theta_6$ (see (\ref{fullmeas})). Then 
we make use of the Grassmann and harmonic delta functions 
$\delta_{56}\delta^{(2,-2)}(u_5,u_6)$ to do the integrals $\int 
du_6 d^8\theta_6$. The next step is to pull the projector 
$(D^+_1)^4$ from the propagator $\Pi_{15}$ out of the integrals 
(it does not contain any integration variables and it only acts on 
the first propagator). After that the remaining projector 
$(D^+_5)^4$ from $\Pi_{15}$ can be used to restore the Grassmann 
vertex $d^8\theta_5$ (everything else under the integral is 
$D^+_5$-analytic). In this way we can free the Grassmann delta 
function $\delta_{15}$ and then do the integral $\int 
d^8\theta_5$. The resulting expression is (up to a numerical 
factor): 

\begin{eqnarray}
  I_1 &=& -\Pi_{14}\Pi_{32}(D^+_1)^4\int {d^4x_5 d^4x_6 du_5\over (15)^3 x^2_{15}
x^2_{56}}\times   \nonumber\\ 
    &\times&  {(D^+_5)^4(D^+_2)^4\over (52)^3}\left({\delta_{12}\over
x^2_{12}}\right) {(D^+_3)^4(D^+_5)^4\over 
(35)^3}\left({\delta_{31}\over x^2_{36}}\right) 
{(D^+_5)^4(D^+_4)^4\over (54)^3}\left({\delta_{14}\over 
x^2_{64}}\right) \;.\nonumber 
\end{eqnarray}

Here all $D^+$'s contain the same $\theta=\theta_1$ but different 
$u$'s, as indicated by their index. Next we distribute the four 
spinor derivatives $(D^+_1)^4$ over the three propagators and use 
the identities (\ref{relations}) (remember that we are only 
interested in the leading term of the correlator, therefore we set 
all $\theta$'s$=0$). The result is: 

\bea I_1(\theta=0) &=& -{(14)(23)\over x^2_{23}x^2_{14}}\int du_5 
\times \nonumber\\ &\times& \left[ { 4i\pi^2\over x^2_{34}} 
{(14)^2(52)(53)\over (51)(54)} g_3 + {4i\pi^2\over x^2_{34}} 
{(13)^2(52)(54)\over (51)(53)}  g_4 + {4i\pi^2\over x^2_{12}} 
{(12)^2(53)(54)\over (51)(52)}  g_1 \right.\nonumber\\ &+& 
\left.(13)(14) {(52)\over (51)} 2\partial_3\cdot\partial_4f + 
(12)(14) {(53)\over (51)} 2\partial_2\cdot\partial_4f +(12)(13) 
{(54)\over (51)} 2\partial_2\cdot\partial_3f \right] \;. \nonumber 
\eea 

Here 

\be\label{2loop} f(x_1,x_2,x_3,x_4) = \int{d^4x_5 d^4x_6\over 
x^2_{15}x^2_{25}x^2_{36}x^2_{46}x^2_{56}} \ee 

and, e.g., 

\be\label{1loop} g_1(x_2,x_3,x_4) = {x^2_{12}\over 4i\pi^2} 
\Box_2f(x_1,x_2,x_3,x_4) = \int{d^4x_5\over 
x^2_{25}x^2_{35}x^2_{45}} \ee 

are two- and one-loop space-time integrals. The last step is to 
compute the harmonic integral. The way this is done is explained 
by the following example: 

\be
\int du_5 {(52)\over (51)} = \int du_5 {D^{++}_5(5^-2)\over (51)} 
= - \int du_5 (5^-2) D^{++}_5 {1\over (51)} =  - \int du_5 
(5^-2)\delta^{(1,-1)}(5,1) = -(1^-2)\;, \nonumber \ee 

where $(1^-2) \equiv u^{-i}_1u^+_{2i}$ and eq. (\ref{hdistr}) has 
been used. Other useful identities needed to simplify the result 
are, e.g.: 

\be\label{cyclid} (1^-2)(13) = -(23) + (1^-3)(12) \ee 

(based on the cyclic property of the $SU(2)$ traces and on the 
defining property $(11^-) = 1$, see (\ref{harco})) and 

\be
(\partial_1+\partial_2)^2 f = (\partial_3+\partial_4)^2 f \quad 
\Rightarrow \quad 2\partial_1\cdot\partial_2 f = 
2\partial_3\cdot\partial_4 f - {4i\pi^2\over x^2_{12}}(g_1+g_2) + 
{4i\pi^2\over x^2_{34}}(g_3+g_4) \ee 

(based on the translational invariance of $f$). So, the end result 
for the first graph in Figure 2 is: 

\bea I_1(\theta=0) &=&  {(14)(23)\over x^2_{14} x^2_{23}}\left\{ 
(21^-)(13)(14) {4i\pi^2 g_2\over x^2_{12}} + (12^-)(23)(24) 
{4i\pi^2 g_1\over x^2_{12}} \right. \nonumber\\ && - 
(23^-)(13)(34) {4i\pi^2 g_4\over x^2_{34}} + (24^-)(14)(34) 
{4i\pi^2 g_3\over x^2_{34}}\label{interm}\\ && + 
\left.(14)(23)2\partial_1\cdot\partial_2 f(1,2,3,4) - 
(12)(34)\left[2\partial_2\cdot\partial_3 f(1,2,3,4) + {4i\pi^2 
g_1\over x^2_{12}}\right]\right\}\;. \nonumber \eea 

The second graph in Figure 5 is obtained by exchanging points 1 
and 3: 

\be
I_2 = I_1(3,2,1,4)\;. \ee 

An important remark concerning these intermediate results is the 
fact that they do not satisfy the harmonic analyticity condition 
(\ref{emhyp}), as one would expect from the property of the free 
on-shell hypermultiplets. Indeed, several terms in (\ref{interm}) 
contain negative-charged harmonics which are not annihilated by 
$D^{++}$. As we shall see below, this important property of 
harmonic analyticity is only achieved after summing up all the 
relevant two-loop graphs. So, let us move on to the topology (b) 
in Figure 4. There are four such graphs shown in Figure 6: 

\vskip 0.5 in 
\begin{center}
\begin{picture}(38000,8000)(0,-4000)

\drawline\fermion[\E\REG](0,0)[6000] \global\advance\pfrontx by 
400 \global\advance\pfronty by 200 \put(\pfrontx,\pfronty){4} 
\global\advance\pmidx by -400 \global\Yone=-1500 
\put(\pmidx,\Yone){$J_1$} 
\drawline\fermion[\N\REG](\pbackx,\pbacky)[6000] 
\global\advance\pfrontx by 400 \global\advance\pfronty by 200 
\put(\pfrontx,\pfronty){3} 
\drawline\fermion[\W\REG](\pbackx,\pbacky)[6000] 
\global\advance\pfrontx by 400 \global\advance\pfronty by 200 
\put(\pfrontx,\pfronty){2} 
\drawline\fermion[\S\REG](\pbackx,\pbacky)[6000] 
\global\advance\pfrontx by 400 \global\advance\pfronty by 200 
\put(\pfrontx,\pfronty){1} \global\Xtwo=\pmidx 
\global\advance\pfronty by -200 \global\Ytwo=\pfronty 
\startphantom \drawline\gluon[\NE\FLIPPED](\pmidx,\pmidy)[3] 
\stopphantom \global\Ythree=\gluonlengthy \global\negate\Ythree 
\global\advance\Ytwo by \Ythree 
\drawline\gluon[\NE\FLIPPED](\Xtwo,\Ytwo)[3] 

\drawline\fermion[\E\REG](10000,0)[6000] \global\advance\pfrontx 
by 400 \global\advance\pfronty by 200 \put(\pfrontx,\pfronty){4} 
\global\advance\pmidx by -400 \put(\pmidx,\Yone){$J_2$} 
\drawline\fermion[\N\REG](\pbackx,\pbacky)[6000] 
\global\advance\pfrontx by 400 \global\advance\pfronty by 200 
\put(\pfrontx,\pfronty){3} 
\drawline\fermion[\W\REG](\pbackx,\pbacky)[6000] 
\global\advance\pfrontx by 400 \global\advance\pfronty by 200 
\put(\pfrontx,\pfronty){2} \global\advance\pfrontx by -400 
\global\Xtwo=\pfrontx \global\Ytwo=\pmidy \startphantom 
\drawline\gluon[\SE\FLIPPED](\pmidx,\pmidy)[3] \stopphantom 
\global\Xthree=\gluonlengthx \global\negate\Xthree 
\global\advance\Xtwo by \Xthree 
\drawline\gluon[\SE\FLIPPED](\Xtwo,\Ytwo)[3] 
\drawline\fermion[\S\REG](\fermionbackx,\fermionbacky)[6000] 
\global\advance\pfrontx by 400 \global\advance\pfronty by 200 
\put(\pfrontx,\pfronty){1} 

\drawline\fermion[\E\REG](20000,0)[6000] \global\advance\pfrontx 
by 400 \global\advance\pfronty by 200 \put(\pfrontx,\pfronty){4} 
\global\advance\pmidx by -400 \put(\pmidx,\Yone){$J_3$} 
\drawline\fermion[\N\REG](\pbackx,\pbacky)[6000] 
\global\advance\pfrontx by 400 \global\advance\pfronty by 200 
\put(\pfrontx,\pfronty){3} \global\Xtwo=\pmidx 
\global\advance\pfronty by -200 \global\Ytwo=\pfronty 
\startphantom \drawline\gluon[\SW\FLIPPED](\pmidx,\pmidy)[3] 
\stopphantom \global\Ythree=\gluonlengthy \global\negate\Ythree 
\global\advance\Ytwo by \Ythree 
\drawline\gluon[\SW\FLIPPED](\Xtwo,\Ytwo)[3] 
\drawline\fermion[\W\REG](\fermionbackx,\fermionbacky)[6000] 
\global\advance\pfrontx by 400 \global\advance\pfronty by 200 
\put(\pfrontx,\pfronty){2} 
\drawline\fermion[\S\REG](\pbackx,\pbacky)[6000] 
\global\advance\pfrontx by 400 \global\advance\pfronty by 200 
\put(\pfrontx,\pfronty){1} 

\drawline\fermion[\E\REG](30000,0)[6000] \global\advance\pfrontx 
by 400 \global\advance\pfronty by 200 \put(\pfrontx,\pfronty){4} 
\global\advance\pmidx by -400 \put(\pmidx,\Yone){$J_4$} 
\global\advance\pfrontx by -400 \global\Xtwo=\pfrontx 
\global\Ytwo=\pmidy \startphantom 
\drawline\gluon[\NW\FLIPPED](\pmidx,\pmidy)[3] \stopphantom 
\global\Xthree=\gluonlengthx \global\negate\Xthree 
\global\advance\Xtwo by \Xthree 
\drawline\gluon[\NW\FLIPPED](\Xtwo,\Ytwo)[3] 
\drawline\fermion[\N\REG](\fermionbackx,\fermionbacky)[6000] 
\global\advance\pfrontx by 400 \global\advance\pfronty by 200 
\put(\pfrontx,\pfronty){3} 
\drawline\fermion[\W\REG](\pbackx,\pbacky)[6000] 
\global\advance\pfrontx by 400 \global\advance\pfronty by 200 
\put(\pfrontx,\pfronty){2} 
\drawline\fermion[\S\REG](\pbackx,\pbacky)[6000] 
\global\advance\pfrontx by 400 \global\advance\pfronty by 200 
\put(\pfrontx,\pfronty){1} 

\put(15800,-3000){Figure 6} 
\end{picture}
\end{center}

The calculation is very similar to the one above, so we just give 
the end result: 

\be
J_1(\theta=0) = - {(23)(34)\over x^2_{23}x^2_{34}}\left[ 
(42^-)(12)^2 {4i\pi^2 g_3\over x^2_{12}} + (24^-)(14)^2 {4i\pi^2 
g_3\over x^2_{14}} + (12)(14) 2\partial_2\cdot\partial_4 
f(1,2,1,4)\right]\;.\nonumber \ee 

Notice the appearance of the two-loop integral $f$ (\ref{2loop}) 
with two points identified, $x_1=x_3$. The other three graphs 
$J_{2,3,4}$ in Figure 3 are obtained by cyclic permutation. 
Finally, putting everything together, using cyclic harmonic 
identities of the type (\ref{cyclid}) as well as the identity (see 
the Appendix) 

\be
\Box_1f(1,2,1,3) = {x^2_{23}\over x^2_{12}x^2_{13}}4i\pi^2 g_4\;, 
\ee 

we arrive at the following final result: 

\begin{eqnarray}\label{answer}
  &&\langle \widetilde q^+\widetilde q^+\vert q^+q^+\vert
  \widetilde q^+\widetilde q^+\vert
q^+q^+\rangle  \nonumber \\ 
  &&\qquad = I_1 + I_2 + J_1 + J_2 + J_3 + J_4  \nonumber \\
  &&\qquad = \left[ (14)^2(23)^2 A_1 + (12)^2(34)^2 A_2 +
(12)(23)(34)(41) A_3\right]\;, 
\end{eqnarray}

where 

\be\label{A12} A_1={(\partial_1+\partial_2)^2 f(1,2,3,4)\over 
x^2_{14}x^2_{23}}\;, \qquad A_2 ={(\partial_1+\partial_4)^2 
f(1,4,2,3)\over x^2_{12}x^2_{34}}\;, \ee 

$$
A_3 = 4i\pi^2 {(x^2_{24} - x^2_{12} -x^2_{14}) g_3 + (x^2_{13} - 
x^2_{12} -x^2_{23}) g_4 + (x^2_{24} - x^2_{23} -x^2_{34}) g_1 + 
(x^2_{13} - x^2_{14} -x^2_{34}) g_2 \over 
x^2_{12}x^2_{14}x^2_{23}x^2_{34}} 
$$

\be\label{A3} 
 + {(\partial_2+\partial_3)^2 f(1,2,3,4)\over x^2_{14}x^2_{23}} +
{(\partial_1+\partial_2)^2 f(1,4,2,3)\over x^2_{12}x^2_{34}}\;. 
\ee 

As pointed out earlier, this result is manifestly harmonic 
analytic (there are only positive-charged harmonics in 
(\ref{answer})). It is also easy to see that the correlator is 
symmetric under the permutations $1\leftrightarrow 3$ or 
$2\leftrightarrow 4$ corresponding to exchanging the two 
$\widetilde q^+\widetilde q^+$ or $q^+q^+$ vertices. 

Finally, we turn to the two other correlators (\ref{3corr}) and 
(\ref{2corr}). The difference in the graph structure is the change 
of flow along some of the hypermultiplet lines. By examining the 
graphs in Figures 5 and 6 it is easy to see that this amounts to 
an overall change of sign in the case (\ref{3corr}) (due to an odd 
number of reversals of $q^+$ propagators and SYM-to-matter 
vertices) or to no change at all in the case (\ref{2corr}) (an 
even number of reversals). Then one has to take into account the 
different symmetry of the new configurations which means that the 
three harmonic structures in (\ref{answer}) have to be symmetrised 
accordingly. This is done with the help of cyclic identities like 

\begin{equation}\label{cycid}
  (14)(23) \ {\stackrel{3\leftrightarrow 4}{\rightarrow}}\ (13)(24) = (12)(34)
  + (14)(23)\;.
\end{equation}

\section{Discussion of the results and asymptotic behaviour}

An interesting technical feature of our calculation is that the 
space-time integrals resulting from the Grassmann and harmonic 
integrations can be written in terms of a second-order 
differential operator acting on the basic scalar two-loop 
integral. Ordinarily, this type of gauge theory calculation will 
produce a set of tensor integrals, which one would first need to 
reduce to scalar integrals algebraically, using algorithms such as 
in \cite{tarasov}. The reason for this unusual property can be 
traced back to an alternative form for the $q^+$ propagator which 
is obtained as follows. Given two points in $x$ space, $x_{1}$ and 
$x_2$, one can define the supersymmetry-invariant difference 

\begin{equation}\label{6.5}
  \hat x_{12} =x_{12} + {2i\over (12)} \left[(1^-2) \theta^+_1 \bar\theta^+_1 -
  (12^-) \theta^+_2 \bar\theta^+_2 + \theta^+_1 \bar\theta^+_2 + \theta^+_2
  \bar\theta^+_1\right] \ .
\end{equation}

Note the manifest G-analyticity of this expression with respect to 
both arguments. Now, with the help of (\ref{6.5}) one can rewrite 
the propagator (\ref{qprop}) in the equivalent form 

\begin{equation}\label{6.12}
  \langle \widetilde q^{+}(1) q^+(2)\rangle =
  {(12)\over \hat x^2_{12}}\ .
\end{equation}

One sees that the whole $\theta$ dependence of the propagator is 
concentrated in the shift (\ref{6.5}). Thus, doing the $\theta$ 
integrals in the above graph calculation effectively amounts to 
taking a couple of terms in the Taylor expansion of the scalar 
propagators. This explains the general structure of the resulting 
space-time integrals. 

We now discuss the explicit space-time dependence of the 
correlation functions. The basic integral we encounter is the 
two-loop one (\ref{2loop}). In principle, it could be obtained by 
Fourier transformation from the known result for the momentum 
space double box (see eq. (\ref{D2}) in the Appendix). 
Unfortunately, this appears to be a very difficult job. It is more 
useful to note that, rewritten as a momentum space diagram, the 
same integral is identical with the ``diagonal box'' diagram shown 
in Figure 7. 

\begin{center}
\begin{picture}(38000,13300)(-4000,-6000)
\put (17000,3500) {\line(0,-1){7000}} \put (17000,0) 
{\circle{10000}} \put (15000,0)  {\line(-1,0){1000}} \put 
(19000,0) {\line(1,0){1000}} \put (16500,4500) {$x_{41}$} \put 
(16500,-4500) {$x_{23}$} \put (13100,-1200) {$x_{12}$} \put 
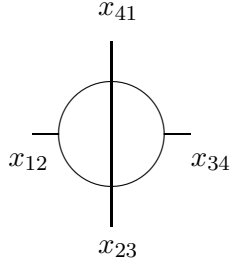
(20000,-1200) {$x_{34}$} \put (10000,-6400) {Figure 7: Diagonal 
box diagram.} 
\end{picture}
\end{center}

This diagram has not yet been calculated for the general off-shell 
case. However in the special case where either $x_{23}=0$ or 
$x_{41}=0$ it is known to be expressible in terms of the function 
$\Phi^{(2)}$ defined in eq. (\ref{Phiexplicit}) in the Appendix. 
For example, for $x_{41}=0$ one has \cite{davuss2loop3point} 

\bear f(x_1,x_2,x_3,x_1) &=& {(i\pi^2)^2\over x_{23}^2} \Phi^{(2)} 
\Bigl({x_{12}^2\over x_{23}^2}, {x_{13}^2\over x_{23}^2}\Bigr)\;. 
\label{fspecial} \ear 

In the same way, the one-loop integral $g$ can, by eq. 
(\ref{C1=Phi1}), be expressed in terms of another function 
$\Phi^{(1)}$ defined in eq. (\ref{Phi1explicit}) in the Appendix: 

\bear g(x_1,x_2,x_3) &\equiv& \int {dx_4 \over 
x_{14}^2x_{24}^2x_{34}^2} = - {i\pi^2\over x_{12}^2} \Phi^{(1)} 
\Bigl( {x_{23}^2\over x_{12}^2} , {x_{31}^2\over x_{12}^2} 
\Bigr)\;. \label{calcg3} \ear 

A further explicit function can be found in the case of some 
particular combination of derivatives on the two-loop integral 
$f$. By exploiting the translation invariance of the 
$x_5$-subintegral we can do the following manipulation on the 
integral, e.g.: 

\bear (\partial_1 +\partial_2)^2 f(1,2,3,4) &=& \int {dx_6 \over 
x_{36}^2 x_{46}^2 } (\partial_1 +\partial_2)^2 \int {dx_5 \over 
x_{15}^2 x_{25}^2 x_{56}^2 } \non\\ &=& \int {dx_6 \over x_{36}^2 
x_{46}^2 } 
\partial_6^2
\int {dx_5 \over x_{15}^2 x_{25}^2 x_{56}^2 } \non\\ &=& 4i\pi^2 
\int {dx_6 \over x_{36}^2 x_{46}^2 } \int {dx_5 \over x_{15}^2 
x_{25}^2 } \delta(x_{56}) \non\\ &=& 
4i\pi^2 
\int {dx_5 \over 
x_{15}^2x_{25}^2x_{35}^2x_{45}^2 }\;. \label{calcA1} \ear 

This 4-point one-loop function is given by 

\bear h(x_1,x_2,x_3,x_4) &\equiv& \int {dx_5 \over 
x_{15}^2x_{25}^2x_{35}^2x_{45}^2 } 
=
-
{i\pi^2\over x_{13}^2x_{24}^2} \Phi^{(1)} \Bigl( 
{x_{12}^2x_{34}^2\over x_{13}^2x_{24}^2} , {x_{23}^2x_{41}^2\over 
x_{13}^2x_{24}^2} \Bigr)\;. \label{calcg4} \ear\no 

Unfortunately, no such trick exists in the case of the combination 
of derivatives, e.g., $(\partial_1 +\partial_3)^2 f(1,2,3,4)$ and 
we do not know the corresponding explicit function. This technical 
problem prevents us from demonstrating the manifest conformal 
invariance of the result. Indeed, if the integrals appearing in 
the coefficients $A_1$ and $A_2$ (\ref{A12}) can be reduced to the 
form (\ref{calcg4}) where the explicit dependence on the two 
conformal cross-ratios is visible, the same is not obvious for the 
third coefficient $A_3$ (\ref{A3}). The property 
eq.(\ref{fspecial}) indicates that $f$ itself has not the form of 
a function of the conformal cross ratios times propagator factors. 
On the other hand, without further information on $f$ we cannot 
completely exclude the possibility that this particular 
combination of derivatives of $f$ also breaks down to conformally 
invariant one-loop quantities, only in a less obvious way than it 
happens for the other ones. 

The only qualitative information about the correlation functions 
we can obtain concerns their asymptotic behaviour when two points 
approach each other. This is done in several steps. Firstly, eq. 
(\ref{Phi1asympt}) gives us information on the coincidence limits 
of $g$, e.g., for $x_3\to x_1$ one has 

\bear g(x_1,x_2,x_3) \quad {\stackrel{\sy x_{31} \to 0}{\sim}} 
\,\, {i\pi^2\over x_{12}^2} \ln x_{31}^2\; . \label{g3asympt} 
\ear\no 

Similarly, for the function $h$ we find: 

\bear h(x_1,x_2,x_3,x_4) \quad {\stackrel{\sy x_{41} \to 0}{\sim}} 
\,\, {i\pi^2\over x_{13}^2x_{21}^2} \ln x_{14}^2\;. 
\label{g4asympt} \ear\no 

This then allows us to determine the asymptotic behaviour of $A_1$ 
and $A_2$: 

\bear A_1 \quad {\stackrel{\sy x_{41} \to 0}{\sim}} \,\, 
-
4\pi^4 {\ln x_{14}^2 \over x_{14}^2x_{23}^2x_{13}^2x_{12}^2 }\;; 
\label{A1asympt} \ear\no

\bear A_2 &=& 4i\pi^2 \, { h(x_1,x_2,x_3,x_4) \over 
x_{12}^2x_{34}^2 } \quad {\stackrel{\sy x_{41} \to 0}{\sim}} \,\, 
-4\pi^4 {\ln x_{14}^2 \over x_{13}^4 x_{12}^4 }\;. 
\label{A2asympt} \ear\no 

The case of $A_3$ requires more work, since the derivatives of $f$ 
appearing here cannot be used to get rid of one integration. 
However, in the case of $(\partial_2 +\partial_3)^2f(1,2,3,4)$ the 
limit $x_4\to x_1$ is finite. We can therefore take this limit 
before differentiation. By a similar argument as above one can 
show that 

\bear (\partial_2 +\partial_3)^2 f(1,2,3,1) &=& 4i\pi^2 
{x_{23}^2\over x_{12}^2x_{13}^2} g_4 \label{A3trick} \ear (this 
identity can also be derived using eq. (\ref{fspecial}) and 
differentiating under the integral in eq. (\ref{C2})). Thus we 
find 

\bear { (\partial_2 +\partial_3)^2f(1,2,3,4) \over 
x_{14}^2x_{23}^2 } \quad {\stackrel{\sy x_{41} \to 0}{\sim}} \,\, 
4i\pi^2 {g_4\over x_{14}^2x_{12}^2x_{13}^2 } \label{A3easyterm} 
\ear \no 

This term is a pure pole term, without logarithmic corrections. 
The same procedure does not work for the last term in $A_3$, since 
here the limit is divergent. We evaluate this term by first 
symmetrising and then differentiating under the integral: 

\bear (\partial_1+\partial_2)^2f(1,4,2,3) &=& \half \Bigl[ 
(\partial_1+\partial_2)^2 + (\partial_3+\partial_4)^2 \Bigr] 
f(1,4,2,3) \non\\ &=& 2i\pi^2 \Bigl[ {g_4\over x_{14}^2} + 
{g_3\over x_{23}^2} + {g_2\over x_{23}^2} + {g_1\over x_{14}^2} 
\Bigr] \non\\ &&\!\!\!\!\!\!\!\! + 4 \biggl\lbrace \int { dx_5dx_6 
\,\,\, x_{15}\cdot x_{26} \over x_{45}^2 x_{36}^2 x_{56}^2 
x_{15}^4 x_{26}^4 } + \int { dx_5dx_6 \,\,\, x_{36}\cdot x_{45} 
\over x_{26}^2 x_{15}^2 x_{56}^2 x_{36}^4 x_{45}^4 } 
\biggr\rbrace\;. \non\\ \label{A3diff1} \ear\no 

The remaining integrals are still singular in the limit $x_4\to 
x_1$ but do not contribute to the ${1\over x_{14}^2}$ - pole. 
After combining the terms involving $g_4$ and $g_1$ with eq. 
(\ref{A3easyterm}) and the explicit $g_i$ - terms appearing in 
$A_3$ one finds that the leading ${1\over x_{14}^2}$ - pole 
cancels out, leaving a subleading logarithmic singularity for 
$A_3$.

\section{Conclusions}

We have seen that both the $A_1$ and $A_2$ terms appearing in the 
four-point function calculations turn out to be reducible to 
one-loop quantities by various manipulations; they can be 
evaluated explicitly and are clearly conformally invariant. 
However, we have not succeeded in reducing the $A_3$ term to a 
one-loop form (although it is not ruled out that such a reduction 
may be possible) and it is consequently more difficult to verify 
conformal invariance for this term since the function 
$f(x_1,x_2,x_3,x_4)$ is not known explicitly. 

Even though $A_3$ is not known explicitly as a function, we have 
seen that it is possible to evaluate its leading behaviour in the 
coincidence limit $x_{14}\sim 0$. The three different tensor 
structures in the correlation function have different 
singularities with the strongest being the one with the known 
coefficient $A_1$. This is given by $(x^2)^{-1}\ln x^2$. In 
section 2 we have shown how one may calculate the leading term (in 
a $\theta$ expansion) of the $N=4$ supercurrent four-point 
function from $A_1$, $A_2$ and $A_3$, and in particular, that we 
can calculate the leading term of the $N=2$ four-point function 
involving two $W^2$ operators and two $\bar W^2$ operators. The 
leading behaviour of the $\theta$-independent term of this 
four-point function in the coincidence limit is therefore 
determined by the leading behaviour of $A_1$. If this were to 
remain true for the higher-order terms in the $\theta$-expansion 
then, since $A_1$ is a known function of invariants, we can use 
the argument of section 3 to compute the asymptotic behaviour of 
four-point functions of $F^2$ and $F\tilde F$ and this would put 
us in a better position to make a comparison with the SG 
computations when they are complete. This point is currently under 
investigation. 

It is interesting to note that some of the qualitative features we 
have found here, such as one-loop box integrals and logarithmic 
asymptotic behaviour, have also been found in instanton 
calculations \cite{inst}. 

In the case of three-point functions it is believed that the 
corrections to the free term in perturbation theory cancel, at 
least at leading order in $1/N_c$. This is certainly not the case 
for four-point functions, and it is not clear precisely what 
relation the perturbative results reported on here should have to 
SG computations. It would be interesting to see what happens at 
three loops, and whether one gets similar asymptotic behaviour in 
the coincidence limit. 

Finally, we note that the four-point functions computed here 
exhibit harmonic analyticity even though the underlying fields, 
the hypermultiplet and $N=2$ SYM gauge field in harmonic 
superspace, are only Grassmann analytic. This is a more stringent 
check of the analyticity postulate of the $N=4$ harmonic 
superspace formalism than the previous three-point check and 
obtaining a positive result on this point is encouraging.

\vspace{20pt} {\bf Acknowledgements:} We would like to thank A. 
Davydychev for useful information concerning refs. 
\cite{davussladder,davuss2loop3point}. This work was supported in 
part by the EU network on Integrability, non-perturbative effects, 
and symmetry in quantum field theory (FMRX-CT96-0012) and by the 
British-French scientific programme Alliance (project 98074). 
\vskip 10pt {\bf Note added:} In a recent e-print \cite{last}, a 
special case of the amplitude considered here has been calculated 
using $N=1$ superspace Feynman rules. Their result corresponds to 
our term $A_2$. \vskip 10pt {\bf Note added in proof:} We would 
like to emphasise that the three components of the amplitude 
$A_1,A_2,A_3$ are {\sl not} trivially related to each other by the 
$SU(4)$ invariance of the $N=4$ theory (see Section 2), as has 
been erroneously assumed in the first reference in \cite{inst}.

\begin{appendix}

\section{Calculation of massless $x$-space diagrams}

\renewcommand{\theequation}{A.\arabic{equation}}
\setcounter{equation}{0} 

The $x$-space Feynman graphs considered here are convergent for 
generic arguments, so that dimensional regularization is never 
needed. We can thus use the fact that in strictly four dimensions 
the (massless) $x$-space and momentum space propagators are 
identical, 

$${1\over p^2 + i \epsilon} \leftrightarrow {1\over x^2 - i\epsilon}$$

This fact allows us to formally rewrite our integrals as momentum 
space integrals, and to draw on results which are available for 
momentum space diagrams (the difference in the treatment of the 
propagator pole will cause some sign changes in transferring the 
formulas below to $x$-space). At the one-loop level, the massless 
3-and 4-point functions have been known for a long time 
\cite{karneu,thovel}. At two loops, a number of exact results for 
3-and 4-point diagrams were obtained by Davydychev and Ussyukina 
\cite{davussladder,davuss2loop3point}. 

Defining the one-loop 3-point Feynman integral with ingoing 
momenta $p_1,p_2,p_3$ (such that $p_1+p_2+p_3 =0$) in $4$ 
dimensions as 

\bear C^{(1)}(p_1^2,p_2^2,p_3^2) 
=
\int {d^4q \over q^2{(q+p_1)}^2{(q+p_1+p_2)}^2 } \label{defC1} 
\ear\no 

one has 

\bear C^{(1)}(p_1^2,p_2^2,p_3^2) = {i\pi^2\over p_3^2} \Phi^{(1)} 
(x,y) \label{C1=Phi1} \ear\no 

Here the dimensionless variables $x,y$ are defined by 

\bear x\equiv {p_1^2\over p_3^2} , \qquad y\equiv {p_2^2\over 
p_3^2} \label{defxy} \ear\no 

The function $\Phi^{(1)}(x,y)$ can be represented explicitly in 
terms of dilogarithms \cite{davussladder}, 

\bear \Phi^{(1)}(x,y) 
=
{1\over \lambda} \Biggl\lbrace 2\Bigl({\rm Li}_2(-\rho x) + {\rm 
Li}_2(-\rho y)\Bigr) +\ln {y\over x} \ln {{1+\rho y}\over {1+\rho 
x}} + \ln (\rho x)\ln (\rho y) + {\pi^2\over 3} \Biggr\rbrace 
\non\\ \label{Phi1explicit} \ear\no 

where 

\bear \lambda(x,y)\equiv \sqrt{(1-x-y)^2-4xy}, \qquad 
\rho(x,y)\equiv 2(1-x-y+\lambda)^{-1} \label{deflambdarho} \ear\no 

(here we assume $\lambda^2 > 0$; the case $\lambda^2 < 0$ requires 
an appropriate analytic continuation). Also the following 
parameter integral representation for $\Phi^{(1)}$ is useful 
\cite{davussladder}, 

\bear \Phi^{(1)}(x,y) = - \int_0^1 {d\xi\over {y\xi^2 + (1-x-y)\xi 
+x}} \Bigl( \ln {y\over x} + 2\ln \xi \Bigr) \label{Phi1parint} 
\ear\no 

from which one can easily read off its asymptotic behaviour  for 
$p_2\to 0$, 

\bear \Phi^{(1)}(x,y)\quad {\stackrel{\sy p_2 \to 0}{\sim}} \,\, - 
\ln p_2^2 \label{Phi1asympt} \ear\no 

The 4-point function 

\bear D^{(1)}(p_1^2,p_2^2,p_3^2,p_4^2,s,t) = \int {d^4 q \over 
q^2(q+p_1)^2(q+p_1+p_2)^2(q+p_1+p_2+p_3)^2 } \label{defD1} \ear 
\no 

($s\equiv {(p_1+p_2)}^2,t\equiv {(p_2+p_3)}^2$) can be expressed 
in terms of the same function as \cite{davussladder} 

\bear D^{(1)}(p_1^2,p_2^2,p_3^2,p_4^2,s,t) 
=
{i\pi^2\over st} \Phi^{(1)}(X,Y) \label{D=phi} \ear\no 

where now 

\bear X\equiv {p_1^2p_3^2\over st}, \qquad Y\equiv 
{p_2^2p_4^2\over st} \label{defXY} \ear\no 

Very similar results were obtained for the basic two - loop 
diagrams depicted in Figure 8 
 \cite{davussladder,davuss2loop3point}.

\begin{picture}(38000,12000)(-4000,-6000)
\put(12000,0){\line(-1,0){1000}} \put(12000,0){\line(2,1){3200}} 
\put(12000,0){\line(2,-1){3200}} 
\put(13200,600){\line(0,-1){1200}} 
\put(14400,1200){\line(0,-1){2400}} \put(9500,0){$p_3$} 
\put(16000,1700){$p_1$} \put(16000,-1700){$p_2$} 
\put(25000,1500){\line(0,-1){3000}} 
\put(26300,1500){\line(0,-1){3000}} 
\put(27600,1500){\line(0,-1){3000}} 
\put(24000,1500){\line(1,0){4600}} 
\put(24000,-1500){\line(1,0){4600}} \put(22500,1500){$p_2$} 
\put(29600,1500){$p_3$} \put(22500,-2500){$p_1$} 
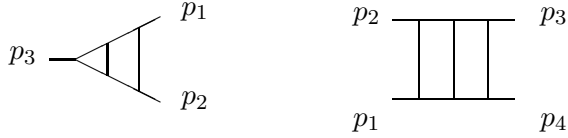
\put(29600,-2500){$p_4$} \put(4000,-5000){Figure 8: a) 2-loop 
triangle diagram, \hspace{5pt} b) 2-loop box diagram.} 
\end{picture}

\noindent The standard two-loop triangle diagram Fig. 8a yields 

\bear C^{(2)}(p_1^2,p_2^2,p_3^2) &=& \int {d^4 q \over 
q^2(q+p_1)^2(q+p_1+p_2)^2} C^{(1)}(q^2,(q+p_1+p_2)^2,p_3^2) \non\\ 
&=& {\biggl( {i\pi^2\over p_3^2} \biggr)}^2 \Phi^{(2)}(x,y) 
\label{C2} \ear\no 

with $x,y$ as before, and 

\bear \Phi^{(2)}(x,y) &=& {1\over \lambda} \Biggl\lbrace 6\Bigl( 
{\rm Li}_4 (-\rho x) + {\rm Li}_4 (-\rho y) \Bigr) + 3\ln {y\over 
x} \Bigl( {\rm Li}_3 (-\rho x) - {\rm Li}_3 (-\rho y) \Bigr) 
\non\\ && + \half \ln^2 {y\over x} \Bigl( {\rm Li}_2 (-\rho x) + 
{\rm Li}_2 (-\rho y) \Bigr) +\fourth \ln^2 (\rho x)\ln^2(\rho y) 
\non\\ && +\half \pi^2 \ln (\rho x)\ln (\rho y) + {1\over 12} 
\pi^2\ln^2 {y\over x} + {7\over 60} \pi^4 \Biggr\rbrace 
\label{Phiexplicit} \ear\no 

Here the polylogarithms are defined as 

\bear {\rm Li}_N (z) &=& {(-1)^N\over (N-1)!} \int_0^1 d\xi 
{\ln^{N-1}\xi\over \xi - z^{-1}} \label{defpolylog} \ear\no 

The momentum-space double box diagram Figure. 8b can be expressed 
in terms of the same function $\Phi^{(2)}$, 

\bear D^{(2)}(p_1^2,p_2^2,p_3^2,p_4^2,s,t) &=& t {\biggl( 
{i\pi^2\over st} \biggr)}^2 \Phi^{(2)}(X,Y) \label{D2} \ear\no

\end{appendix}

\end{document}